\pdfoutput=1        
\documentclass[sigplan,10pt,nonacm]{acmart}         
\makeatletter                   
\def\mdseries@tt{m}             
\makeatother        

\makeatletter
\renewcommand\@formatdoi[1]{\ignorespaces}
\makeatother

\usepackage[export]{adjustbox}
\usepackage{listings}
    
\usepackage[newfloat=true,frozencache=true]{minted} 
\usepackage{color}                                                                   
\usepackage{xcolor}                                                                  
\usepackage{changepage} 
                                                                                     
\usepackage{url}                                                                     
\hypersetup{hidelinks=false,colorlinks=true,linkcolor=blue,urlcolor=blue,citecolor=blue,anchorcolor=blue} 
\usepackage{breakurl}                                                                

\usepackage{tabularx}
\usepackage{booktabs} 

\usepackage{cleveref}
\crefformat{section}{\S#2#1#3}
\crefformat{subsection}{\S#2#1#3}
\crefformat{subsubsection}{\S#2#1#3}
\crefrangeformat{section}{\S\S#3#1#4 to~#5#2#6}
\crefmultiformat{section}{\S\S#2#1#3}{ and~#2#1#3}{, #2#1#3}{ and~#2#1#3}

\usepackage{microtype}
                                                 
\usepackage[utf8]{inputenc}                                                          

\usepackage[iso,american,cleanlook]{isodate}    

\usepackage{rviewport}                                                               

\usepackage{tikz} 
\usepackage[inline]{enumitem} 
\newcommand*\circled[1]{\tikz[baseline=(char.base)]{
            \node[shape=circle,fill,inner sep=2pt] (char) {\textcolor{white}{#1}};}}

\newcommand\Invisible[1]{                                                            
  \marginpar{\color{white}{\fontsize{.5}{.5}\selectfont #1 }}                        
}

\newcommand{\Exclude}[1]{}

\newcommand\Boldly[1]{\vspace{0.5 \baselineskip} \noindent \textbf{$\blacktriangleright$} \textbf{#1} \noindent}

\definecolor{Gray95}{gray}{0.95}
                                                                                     
\newcommand{\AtFoot}[1]{\let\thefootnote\relax\footnotetext{{#1}}}

\acmConference[EuroSys'19]{European Conference on Computer Systems}{March 26--28, 2019}{Dresden, Germany}
\acmYear{2019}
\acmISBN{} 
\acmDOI{10.1145/nnnnnnn.nnnnnnn}
\startPage{1}

\usepackage{booktabs}   
\usepackage{subcaption} 

\newcommand{\orcidicon}[1]{\href{https://orcid.org/#1}{\includegraphics[scale=0.06]{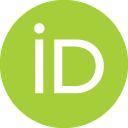}}}

\begin{document}

\title[]{TWA -- Ticket Locks Augmented with a Waiting Array} 


\author{Dave Dice \orcidicon{0000-0001-9164-7747}}
\orcid{0000-0001-9164-7747}             
\affiliation{
  \institution{Oracle Labs}             
}
\email{first.last@oracle.com}            

\author{Alex Kogan \orcidicon{0000-0002-4419-4340}} 
\orcid{0000-0002-4419-4340} 
\affiliation{
  \institution{Oracle Labs}             
}
\email{first.last@oracle.com}          


\begin{abstract}

The classic \emph{ticket lock} \cite{focs79,cacm79-reed,tocs91-MellorCrummey} consists of 
\texttt{ticket} and \texttt{grant} fields.
Arriving threads atomically fetch-and-increment \texttt{ticket} to obtain an assigned ticket value,
and then wait for \texttt{grant} to become equal to that value, at which point the thread holds the lock.
The corresponding unlock operation simply increments \texttt{grant}.   
This simple design has short code paths and fast handover (transfer of ownership) 
under light contention, but may suffer degraded scalability under high contention when multiple 
threads busy wait on the \texttt{grant} field -- so-called \emph{global spinning}.  


We propose a variation on ticket locks where long-term waiting threads -- those with
an assigned \texttt{ticket} value far larger than \texttt{grant} -- wait on locations in a  
\emph{waiting array} instead of busy waiting on the \texttt{grant} field.  
The single waiting array is shared among all locks. 
Short-term waiting is accomplished in the usual manner on the \texttt{grant} field.  
The resulting algorithm, \underline{TWA}, improves on ticket locks by limiting the number of 
threads spinning on the \texttt{grant} field at any given time,
reducing the number of remote caches requiring invalidation from the store that releases the lock.
In turn, this accelerates handover, and since the lock is held throughout the handover operation, 
scalability improves.  
Under light or no contention, TWA yields performance comparable to the classic ticket lock. 
Under high contention, TWA is substantially more scalable than the classic ticket lock, and
provides performance on par or beyond that of scalable queue-based locks such 
as MCS \cite{tocs91-MellorCrummey} by avoiding the complexity and extra accesses incurred 
by the MCS handover operation while also obviating the need for maintaining queue elements.

We provide an empirical evaluation, comparing TWA against ticket locks and MCS for various
user-space applications, and within the Linux kernel.

\Invisible{TWA avoids the complexity and extra accesses incurred by scalable queue-based locks, 
such as MCS the handover path, providing performance above or 
beyond that of MCS at high contention.} 

\Invisible{Relative to MCS we also incur fewer coherence misses during handover.  
Under contention, MCS will generally induce a coherence miss fetching the owner node's \texttt{Next} 
field, and another miss to set the \texttt{flag} field in the successor's node, whereas our 
approach causes only one miss in the critical handover path.}


\end{abstract}




\maketitle

\thispagestyle{fancy}

\Invisible{Oracle ID accession number ORA190196} 
\Invisible{EuroSys 2019 submission number 257} 

\Invisible{
*  Tension; trade-off
*  Primum non-nocere ; bound harm; hippocratic oath
*  Collision probability in Waiting Array is equivalent to "Birthday Paradox".
*  dispersed; diffused; distributed; disseminate; split; fractured; sharded; 
   decomposed; dilute; spread; 
*  deconstruct; decouple; 
*  faithful; realistic; accurate; fidelity; veracity; authentic; adherent; 
*  A short idea expressed in a long paper
*  subsume; subducts
*  performance predictability; variability; variance; consistent; 
*  adhere to principle of least surprise
*  divert; revert; fall-back; fail-over
*  desiderata; performance goals; target; ideal; aspire; 
*  devolve; degenerate; converge; trend toward; tend towards
*  slipstream; 
*  confers
*  affords; encumbrance
*  defensible decision/design
*  supplant
*  consequent; ensuant; ensue; pursuant; arising; by virtue of; 
*  constrain; limit; guard; clamp; guard; cap; protect; restrict; 
*  release; surrender; relinquish; unlock; abjure; 
*  Thought experiment : Gedankenexperiment
*  assiduously avoid
*  relax 
*  Minimize
*  Inter-node miss vs intra-node coherence miss
*  Augment and extent
*  race; inopportune interleaving; intervened; window
*  surpass; exceed
*  Slot; Array element; Table element; 
*  Incremental cost
*  in-effect; prevailing preference; transparent; 
*  Numerous; myriad; plurality; 
*  without surrendering ...; trade-off
*  competitive with
*  suppress bias; inhibit; refrain; squelch;
*  contrived; intentionally configured; artificially; 
*  we observe in passing and without further comment ...
*  favor; at the expense of
*  avoid dilemma
*  MCS : Each arriving thread uses an atomic instruction to append 
   a ``node'' (representing that thread) to tail of a chain of waiting threads, 
   forming an explicit queue, and then busy waits on a field within that node.  
*  MCS : node is proxy for thread
*  MCS : chain ; explicit queue of nodes where node represents waiting thread.
*  MCS is usually considered as an alternative to ticket locks.
*  succession : handoff vs handover
*  QoI = Quality-of-implementation issue
*  SPARC oddities : no 64-bit SWAP; no FAA; emulate all with CAS; MOESI
*  Claim : performance TWA >> Max(MCS, TKT) 
*  TWA unlock path is slightly longer than TKT, but handoff is accomplished first/early.  
*  Ambient; native; free-range; 
*  Supports our claim ...
*  models; reflects; mirrors; 
*  Waiting threads interfere with handover
*  wrap around; rollover; overflow; aliasing; ABA; 
   rollover recurrence could result in progress failure and hang
   waiting thread fails to exii long-term waiting mode
   missed wakeup -- LT thread fails to observe change in WA[x] 
*  Optimization : really "cycle shaving" 
*  Influence of API design on lock : 
   @  pass/convey from lock-to-unlock : pass/convey
      May need to add extra field in lock body to pass
      That field may induce additional coherence traffic
   @  scoped locking; lexically balanced -- allows on-stack queue nodes
   @  express as closure : like java synchronized block
*  Inter-lock hash collisions vs intra-lock collisions
*  Ticket locks made less repulsive
*  overall path length increases; but critical path decreases
*  Cycle shaving game
*  Ticket Locks are \emph{context-free locks}. 
*  latency withing contended CS implies scalability !
*  Readers are not simply passive observers ; Quantum : readers change state
*  Not strictly deterministic; reliable performance
*  Collision : false notification
*  point-to-point 1:1 vs one-to-many communcation
*  Amenable to MONITOR-MWAIT
*  Wait-away; wait-aside
*  dissipate central contention
*  TKTWA5 : invisible readers/waiters
   TKTWA7 : visible readers/waiters
*  Analogy : bakery/deli with TicketAllocator and NowServing
   Crowd/mob around NowServing impedes handover
   MCS is a true queue - line
   TWA : assign waiting place based on ticket
   Move to counter when turn is near
*  Waiting Array values are hints -- advisory
   Can not reason directly about tranfer of ownership from absence or presence of
   values on the waiting array.  Only changes.  
   Waiting thread must always consult ground truth in "grant" field.
*  Ticket lock field names
   Ticket-Grant; Head-Tail; Request-Admit; Arrive-Enter; nextticket-NowServing;
   Next-Owner; Taken-Turn;
*  probabilistic conflict avoidance 
*  Wait-away : distal
*  analogy : crowd at door of bathroom
   larger crowd makes it harder to exit and "pass" ownership
   wait elsewhere
*  ondeck; immediate successor; next-up; 
*  circumstantial correlation vs causation : how to establish etiology
*  Ticket-based form : large performance sensitivity to placement of ticket/grant fields
   sequestration and isolation vs compact colocation 
   Fat vs compact
   
}

\section{Introduction}


The classic ticket lock is compact and has a very simple design.  The acquisition
path requires only one atomic operation -- a fetch-and-add to increment the ticket -- 
and the unlock path requires no atomics.  On Intel systems, fetch-and-add is implemented
via the \texttt{LOCK:XADD} instruction so the \emph{doorway phase} \cite{ArtOf} is wait-free.
Under light or no contention, the handover latency, 
defined as the time between the call to unlock and the time a successor is
enabled to enter the critical section, is low.  Handover time impacts the
scalability as the lock is held throughout handover, increasing the effective
length of the critical section \cite{isca10-eyerman,podc18-aksenov}.  
Typical implementations use 32-bit integers for the \texttt{ticket} and \texttt{grant} 
variable.  Rollover is not a concern as long as the number of concurrently 
waiting threads on a given lock never exceeds $2^{32}-1$ \footnote{
Developers assume this constraint is always satisfied -- having more that $2^{32}-1$ waiting threads is not
considered a practical concern.}. 
A ticket lock is in \emph{unlocked} state when \texttt{ticket} and \texttt{grant} are equal.
Otherwise the lock is held, and the number of waiters is given by \texttt{ticket - grant - 1}.  
Ignoring numeric rollover, \texttt{grant} always lags or is equal to \texttt{ticket}.  
The increment operation in unlock either passes ownership to the immediate successor,
if any, and otherwise sets the state to unlocked.

Ticket locks suffer, however, from a key scalability impediment.  All threads
waiting for a particular lock will busy wait on that lock's \texttt{grant} field.
An unlock operation, when it increments \texttt{grant}, invalidates the 
cache line underlying \texttt{grant} for all remote caches where waiting threads are scheduled. 
In turn, this negatively impacts scalability by retarding the handover step. 
Ticket locks use global spinning, as all waiting threads monitor the central
lock-specific \texttt{grant} variable.   

\begin{figure}[h]                                                                    
\includegraphics[width=8.5cm]{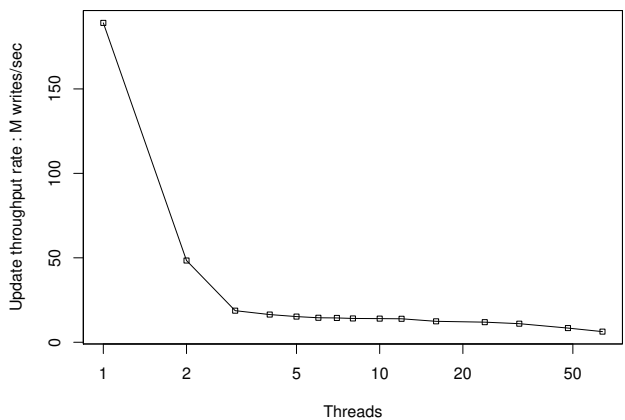} 
\caption{Invalidation Diameter}                                                   
\label{Figure:InvalidationDiameter}                                                                  
\end{figure}

In Figure-\ref{Figure:InvalidationDiameter} we show the impact of readers on a single 
writer.  We refer to the number of participating caches as the \emph{invalidation diameter} \cite{eurosys17-dice}.  
The \texttt{Invalidation Diameter} benchmark spawns $T$ concurrent threads,  with $T$ shown on the
X-axis.  A single writer thread loops, using an atomic fetch-and-add primitive to update a shared 
location.  The other $T-1$ threads are readers. They loop, fetching the value of that location.  
The shared variable is sequestered to avoid false sharing and is the sole occupant of its underlying cache sector. 
We present the throughput rate of the writer on the Y-axis.  
As we increase the number of concurrent readers, the writer's progress is slowed.
This scenario models the situation in ticket locks where multiple waiting threads 
monitor the \texttt{grant} field, which is updated by the current owner during handover.  
The benchmark reports the writer's throughput at the
end of a 10 second measurement interval.  The data exhibited high variance due to the NUMA placement
vagaries of the threads and the home node of the variable.  As such, for each data point show,
we took the median of 100 individual runs, reflecting a realistic set of samples. 
The system-under-test is described in detail in \cref{section:Empirical}.  
\Invisible{As can be seen in the graph, increasing the number of readers degrades the performance of the writer.} 

\Invisible{What we really want to measure store-to-load visibility latency.  
That is, we want to know how multiple readers might affect the lag between
a store and when that store becomes visible to a set of busy-waiting readers. 
This reflects the ticket lock handover operation. 
In theory, for the ``Invalidation Diameter'' experiments, the stores could become visible
quickly to the readers, but the store itself is delayed by the readers, and does not
retire or ``return'' until after some delay.}  

The \emph{MCS lock} \cite{tocs91-MellorCrummey} is the usual alternative to
ticket locks, performing better under high contention, but
also having a more complex path and often lagging behind ticket locks under no or light contention.
In MCS, arriving threads use an atomic operation to append an element,
commonly called a ``queue node'', to the tail of a linked list of waiting threads,
and then busy wait on a field within that element, avoiding global spinning.
The list forms a queue of waiting threads.  
The lock's tail variable is explicit and the head -- the current owner --
is implicit. When the owner releases the lock it reclaims the element it
originally enqueued and sets the flag in the next element, passing ownership.
To convey ownership, the MCS unlock operator must identify the successor, if any, and then
store to the location where the successor busy waits.
The handover path is longer than that of ticket locks and accesses more distinct
shared locations.  MCS uses so-called local waiting where at most one thread is waiting on a given
location at any one time.  As such, an unlock operation will normally
need to invalidate just one location -- the flag where the successor busy waits.
Under contention, the unlock operator must fetch the address of the successor
node from its own element, and then store into the flag in the successor's element,
accessing two distinct cache lines, and incurring a dependent memory access to reach the successor.
In the case of no contention, the unlock operator must use an atomic compare-and-swap
operator to detach the owner's element.

\Invisible{Under classic MCS, arriving threads append an element, commonly called a ``queue node'', 
to the tail of the linked list of waiting threads and then busy wait on a flag within that 
element, avoiding global spinning.
The lock's tail variable is explicit and the head -- the current owner -- is implicit. 
When the owner releases the lock it reclaims the element it originally enqueued and sets 
the flag in the next element, passing ownership.
MCS uses so-called local waiting where at most one thread is waiting on a given
location at any one time.  As such, an unlock operation will normally
need to invalidate just one location -- the flag where the successor busy waits. 
Under contention, the unlock operator must fetch the address of the successor
node from its own element, and then store into the flag in the successor's element, 
accessing two distinct cache lines, and incurring a dependent memory access to reach the successor.
In the case of no contention, the MCS unlock operator must use an atomic compare-and-swap
operator to detach the owner's element.  
Ticket locks and TWA do not require indirection or dependent accesses in the unlock path.} 

One MCS queue node instance is required for each lock a thread currently holds, and
an additional queue node is required while a thread is waiting on a lock.
Queue nodes can not be shared concurrently and can appear on at most one queue
-- be associated with at most one lock -- at a given time.
The standard POSIX \texttt{pthread\_mutex\_lock} and \texttt{pthread\_\allowbreak{}mutex\_\allowbreak{}unlock} 
operators do not require scoped or lexically balanced locking.  
As such, queue nodes can not be allocated on stack.   
Instead, MCS implementations that expose a standard POSIX interface will typically allocate elements 
from thread-local free lists, populated on demand
\footnote{Threads might also \texttt{malloc} and \texttt{free} queue nodes as needed, but 
most malloc allocators are not sufficiently scalable.  Also, many malloc implementations themselves
make use of POSIX locks, resulting in reentry and recursion if a lock implementation were
to try to call \texttt{malloc} which in turn would need to acquire a lock.
We note that the MCS ``K42'' variant \cite{K42,Scott2013} allows queue nodes to 
be allocated on stack -- they are required only while a thread waits -- 
but at the cost of a longer path with more accesses to shared locations.}. 

\Invisible{MCS requires the address of queue node inserted by the owner to 
be passed to the corresponding unlock operator, where it will be used to identify 
a successor, if any.} 

The standard POSIX interface does not provide any means to pass information 
from a lock operation to the corresponding unlock operator.  As such, the address of 
the MCS queue node inserted by the owner thread is usually recorded in the lock 
instance so it can be conveyed to the subsequent unlock operation to identify the successor, if any.
Accesses to the field that records the owner's queue node address may themselves generate 
additional coherence traffic, although some implementations may avoid such accesses to 
shared fields by storing the queue node address in a thread-local associative structure that maps 
lock addresses to the owner's queue node address.

Ticket locks and TWA require no indirection or dependent accesses in the unlock path
and also avoid the need for queue elements and the management thereof.
The queue of waiting threads is implicit in ticket locks and TWA, and explicit in MCS.
MCS, ticket locks and TWA all provide strict FIFO admission order.

Ticket locks are usually a better choice under light or no contention, while MCS
locks are more suitable under heavy contention \cite{Middleware16-Antic,ols12}.
By employing a waiting array for long-term waiting, TWA achieves the best of the two worlds,
as demonstrated by our empirical evaluation with multiple user-space applications and within 
the Linux kernel \footnote{The Linux kernel switched from 
ticket locks to MCS-based locks in 2014 \cite{LWN2014}.}. 

We note in passing that under conditions of intermittent or no contention when no successor
is visible, MCS may require an atomic compare-and-swap operation to detach the onwer's queue
node from the MCS tail pointer.  (When waiting threads are visible in the MCS chain, the MCS unlock
operator accomplish succession with just a store into the next thread's queue node structure).
In contrast, ticket locks never require an atomic read-modify-write instruction in the unlock path.   
We do not believe, however, that the atomic instruction accounts for the difference 
in performance between MCS and ticket locks at low contention.  
For instance if we replace the load-increment-store sequence found in the unlock path of the ticket
lock algorithm with an atomic fetch-and-add, we observe no appreciable change in performance.
Broadly, on modern processors, we find that coherence traffic dictates performance, and that
an atomic compare-and-swap is the perforamnce equivalent of a simple store.  

\Invisible{Oddly, when incurring a coherence miss, fetch-and-add is \emph{faster} than 
\texttt{XADD}; load-increment-store; add-register-to-memory; etc.} 

\Invisible{An additional confounding fact is that under certain loads, the MCS unlock operator
may execute futile CAS operations that generate unnecessary coherence traffic, and that the 
unlock operator may need to busy-wait to allow an arriving successor to update the \texttt{next} pointer
in the owner's qnode.}

\section{The TWA Algorithm} 

TWA builds directly on ticket locks.  We add a new \emph{waiting array} for long-term waiting.
The array is shared amongst all threads and TWA locks in an address space.  
Arriving threads use an atomic fetch-and-increment instruction to advance the \texttt{ticket} value,
yielding the lock request's assigned ticket value,  and then fetch \texttt{grant}.  
If the difference is 0 then we have 
uncontended acquisition and the thread may enter the critical section immediately.  
(This case is sometimes referred to as the lock acquisition \emph{fast-path}).  
Otherwise TWA compares the difference to the \texttt{LongTermThreshold} parameter.  
If the difference exceeds \texttt{LongTermThreshold} then the thread enters the
long-term waiting phase.  Otherwise control proceeds to the short-term waiting phase, which
is identical to that of normal ticket locks; the waiting thread simply waits for \texttt{grant}
to become equal to the ticket value assigned to the thread.  While \texttt{LongTermThreshold} 
is a tunable parameter in our implementation, we found a value of $1$ to be suitable 
for all environments, ensuring that only the immediate successor waits in short-term mode.  
All data reported below uses a value of 1.  

A thread entering the long-term waiting phase first hashes its assigned ticket value to form
an index into the waiting array.  Using this index, it fetches the value from the array and then
rechecks the value of \texttt{grant} \footnote{The recheck step is needed to avoid races between lock and
unlock operations. Specifically, in the window between the load of \texttt{grant} and the recheck,
the owner might have released the lock.  Absent the recheck, the algorithm would be vulnerable to lost wakeups.
Similar \emph{recheck} idioms appear in other constructs, such as the Linux kernel \emph{futex} mechanism, where
waiting threads check a condition, prepare for long-term waiting, and then recheck the condition
before commiting to long-term waiting.}.  
If the observed \texttt{grant} value changed, 
the thread rechecks the difference between that new value and its assigned ticket value, and 
decides once again on short-term versus long-term waiting.  If \texttt{grant} was 
unchanged, the thread then busy waits for the waiting array value to change, at 
which point it reevaluates \texttt{grant}.  When \texttt{grant} is found to be sufficiently
near the assigned ticket value, the thread reverts to normal short-term waiting.
The values found in the waiting array have no particular meaning, except to 
conservatively indicate that a \texttt{grant} value that maps to that index has changed, 
and rechecking of \texttt{grant} is required for waiters on that index.  
As rollover is a concern in the waiting array, we use 64-bit integers, so in practice, 
rollover never occurs. 

The TWA unlock operator increments \texttt{grant} as usual from $U$ to $U+1$ and then 
uses an atomic operator to increment the location in the waiting array that corresponds
to threads waiting on ticket value $U+1+LongTermThreshold$, notifying long-term 
threads, if any, that they should recheck \texttt{grant}.  
An atomic operation is necessary as the location may be subject to hash collisions.  
We observe that this change increases the path length in the unlock operator, but crucially 
the store that effects handover, which is accomplished by a non-atomic increment 
of \texttt{grant}, happens first.  
Given a \texttt{LongTermThreshold} value of 1,
we expect at most one thread, the immediate successor, to be waiting on \texttt{grant}.  
Updating the waiting array occurs \emph{after} handover and outside the critical section.
\Invisible{and does not influence scalability} 

All our experiments use a waiting array with 4096 elements, although
ideally, we believe the waiting array should be sized as a function of the number
of CPUs in the system.  (A similar approach is used to size the \emph{futex} hash
table array in the Linux kernel \footnote{\url{https://blog.stgolabs.net/2014/01/futexes-and-hash-table-collisions.html}}.)  
Hash collisions in the table are benign,
at worst causing unnecessary rechecking of the \texttt{grant} field.  Specifically, collisions are a 
performance and quality-of-implementation concern that does not impact correctness.  
A larger waiting array table will reduce the collisions rate but might increase cache pressure.
We note that the odds of inter-lock collision are equivalent to those given by the 
``Birthday Paradox'' \cite{BirthdayParadox}.  Our hash function is cache-aware and intentionally 
designed to map adjacent ticket values to different 128-byte cache sectors underlying
the waiting array, to reduce false sharing among long-term waiters.  
We multiply the ticket value by 127 and then \texttt{EXCLUSIVE-OR} that result with the address of the lock, 
and then mask with $4096-1$ to form an index into the waiting array.
We selected a small prime $P=127$ to provide the equidistribution
properties of a \textit{Weyl sequence} \cite{jss03-marsaglia} and also to thwart the automatic stride-based
hardware prefetch mechanism which can artifically induce false sharing.
Multiplication by 127 is easily strength-reduced to a shift and subtract.  
We include the lock address into our deterministic hash 
to avoid the situation where two locks might operate in an entrained fashion, with 
ticket and grant values moving in near unison, and thus suffer from excessive 
inter-lock collisions.  A given lock address and ticket value pair always hashes to the same index.
The hash computed in the unlock operator must target the same index as the corresponding hash 
in the long-term waiting path.  We also note that near collisions can result in false sharing,
when two accesses map to distinct words in the same cache sector.  

\Invisible{We claim, that in terms of collisions, 1 lock with 10 waiting threads should
be equivalent to 2 locks with 5 waiting threads each.} 

TWA leaves the structure of the ticket lock unchanged, allowing for easy adoption.  
As the instance size remains the same, the only additional space cost for TWA is 
the waiting array, which is shared over all locks, reflecting a one-time space cost. 

\Invisible{Precautionary increment ... We say ``precautionary'' as there might not be any long-term waiting threads} 

The TWA fast-path for acquisition remains unchanged relative to ticket locks. 
The unlock path adds an increment of the waiting array, to notify
any long-term waiters sufficiently ``near'' the front of the queue that they should transition 
from long-term to short-term waiting.
We note that TWA doesn't reduce overall coherence traffic, but does act to reduce
coherence traffic in the critical handover path, constraining the invalidation
diameter of the store in unlock that accomplishes handover. 
TWA thus captures the desirable performance aspects of both MCS locks 
and ticket locks. 

Listing-\ref{Listing:TWA-py} depicts a pseudo-code implementation of the TWA algorithm.
Lines 7 through 16 reflect the classic ticket lock algorithm and lines 20 through 71 show TWA.
TWA extends the existing ticket lock algorithm by adding lines 41 through 57 for long-term waiting,
and line 71 to notify long-term waiters to shift to classic short-term waiting.  
\Invisible{For the purposes of explication, we assume the compiler avoids reordering
and emits fences as necessary to provide sequential consistency for accesses the
ticket and grant fields, and the waiting array.} 

\subsection{Example Scenario -- TWA in Action}


\begin{enumerate}[label=\protect\circled{\footnotesize\arabic*}, leftmargin=6.5mm]
\item Initially the lock is in \emph{unlocked} state with \texttt{Ticket} and \texttt{Grant} both 0.
\item Thread \emph{T1} arrives at Listing-\ref{Listing:TWA-py} line 34 attempting to acquire the lock.  
\emph{T1} increments \texttt{Ticket} from 0 to 1, and the atomic \texttt{FetchAdd} operator returns the
original value of 0 into the local variable \texttt{tx}, which holds the assigned ticket value for the 
locking request.  
At line 36 \emph{T1} then fetches \texttt{Grant} observing a value of 0. 
Since \texttt{tx} equals that fetched value, we have uncontended lock acquisition.
\emph{T1} now holds the lock and can enter the the critical section immediately, without waiting,
via the fast path at line 39.  
\item Thread \emph{T2} now arrives and tries to acquire the lock.
The \texttt{FetchAdd} operator advances \texttt{Ticket} from 1 to 2 and returns 1, the assigned ticket,
into \texttt{tx} at line 35.  
\emph{T2} fetches \texttt{Grant} and notes that \texttt{tx} differs from that value by 1.
The \texttt{dx} variable holds that computed difference, which reflects the number of threads between
the requester and the head of the logical queue, which is the owner.  
\emph{T2} has encountered contention and must wait.  
The difference is only 1, and \emph{T2} will be the immediate successor,  so \emph{T2} proceeds to line 60 
for short-term waiting similar to that used in classic ticket locks shown at line 10.
\emph{T2} waits for the \texttt{Grant} field to become 1.  
\item Thread \emph{T3} arrives and advances \texttt{Ticket} from 2 to 3, with the \texttt{FetchAdd} operator
returning 2 as the assigned ticket.  The difference between that value (2) and the value of \texttt{Grant}(0) fetched at line 64  
exceeds the \texttt{LongTermThreshold} (1), so \emph{T3} enters the path for long-term waiting at
line 49.  \emph{T3} hashes its observed ticket value of 2 into an index \texttt{at}, say $100$, in the long-term
waiting array and then fetches from \texttt{WaitArray[at]} observing $U$.  To recover from potential races with
threads in the unlock path, \emph{T3} rechecks that the \texttt{Grant} variable remains unchanged (0) at 
line 49 and that the thread should continue with long-term waiting.  Thread \emph{T3} busy waits
at lines 52-53 on the \texttt{WaitArray} value. 
\item Thread \emph{T4} arrives, advances \texttt{Ticket} from 3 to 4, obtaining a value in its \texttt{tx}
variable of 3.  Similar to \emph{T3}, \emph{T4} enters the long-term.  \emph{T4} hashes its assigned ticket value of 3 yielding
an index of, say, $207$, and fetches \texttt{WaitArray[207]} observing $V$. 
\emph{T4} then busy waits, waiting for \texttt{WaitArray[207]} to change from $V$ to any other value. 
\item Thread \emph{T1} now releases the lock, calling \texttt{TWARelease} at line 63.  
\emph{T1} increments \texttt{Grant} from 0 to 1 at line 67, passing ownership to \emph{T2}
and sets local variable \texttt{k} to the new value (1).   
\item Thread \emph{T2} waiting at lines 60-61 notices that \texttt{Grant} changed to match 
its \texttt{tx} value.  \emph{T2} is now the owner and may enter the critical section.
\item Thread \emph{T1}, still in \texttt{TWARelease} at line 71 then hashes $k + LongTermThreshold$ (the sum is 2)
to yield index $100$ and then increments \texttt{WaitArray[100]} from $U$ to $U+1$.  
\item Thread \emph{T3} waiting at lines 52-53 observes that change, rechecks \texttt{Grant}, sees that it
is close to being granted ownership, exits the long-term waiting loop and switches to classic
short-term waiting at lines 60-61.  \emph{T1} has promoted \emph{T3} from long-term to short-term waiting
in anticipation of the next unlock operation, to eventually be performed by \emph{T2}.  
\item Thread \emph{T1} now exits the \texttt{TWARelease} operator. 
\item Thread \emph{T2} is the current owner, thread \emph{T3} is waiting in short-term mode,
and thread \emph{T4} is waiting in long-term mode. 
\end{enumerate}



\begin{listing}[h]         
\begin{adjustwidth}{1em}{0pt}
\inputminted[linenos,bgcolor=Gray95,fontsize=\scriptsize]{py}{excerpt-TWA.py} 
\vspace{-0.5cm} 
\captionof{listing}{Simplified Python-like Implementation of TWA\label{Listing:TWA-py}}
\end{adjustwidth}
\end{listing} 


\section{Related Work}

\Exclude{While mutual exclusion remains an active research topic, we focus on locks
closely related to our design.} 


\citet{tocs91-MellorCrummey} proposed ticket locks with \emph{proportional backoff}. 
Waiting threads compare the value of their ticket against the \texttt{grant} field.
The difference reflects the number of intervening threads waiting.  That value is then multiplied
by some tunable constant, and the thread delays for that period before rechecking \texttt{grant}.
The intention is to reduce futile polling that might induce unnecessary coherence traffic.
The constant is platform- and load-dependent, and requires tuning.  In addition, while the approach
decreases the futile polling rate on \texttt{grant}, and may be used in conjunction with
polite waiting techniques \cite{eurosys17-dice}, it does not decrease the invalidation diameter. 
TWA and ticket locks with proportional backoff both makes a distinction
among waiting threads based on their relative position in the queue.  

Partitioned Ticket Locks \cite{spaa11-dice} augment each ticket lock with a constant-length
private array of \texttt{grant} fields, allowing for \emph{semi-local waiting}.  Critically, the 
array is not shared between locks, and to avoid false sharing within the array, the memory footprint of each
lock instance is significantly increased.  
Ticket Lock ``AWN'' \cite{Ticket-AWN} also uses per-lock array for semi-local waiting.
Anderson's array-based queueing lock \cite{tpds90-Anderson,dc03-Anderson} is also based on ticket locks.
It employs a waiting array for each lock instance, sized to ensure there is at least one array element 
for each potentially waiting thread, yielding a potentially large footprint.  
The maximum number of participating threads must be known in advance when initializing the array.
Such dynamic sizing also makes static allocation of Anderson's locks more difficult than would be the
case for a lock with a fixed size, such as TWA.  

Fu et al. ~\cite{tpds97-fu, tpds98-huang} describe a mutual exclusion scheme that avoids 
global spinning, but the technique has long paths and accesses large numbers of shared variables.  


Various authors \cite{ispan05-ha,Middleware16-Antic} have suggested switching adaptively
between MCS and ticket locks depending on the contention level.  While workable, this adds
considerable algorithmic complexity, particularly for the changeover phase, and requires tuning.  
\citet{asplos94-lim} suggested a more general framework for switching locks at runtime.  
\Invisible{Concerns: reactivity response time; hysteresis and damping; chase; hunt; ring} 

\section{Empirical Evaluation}
\label{section:Empirical} 

Unless otherwise noted, all data was collected on an Oracle X5-2 system.  
The system has 2 sockets, each populated with 
an Intel Xeon E5-2699 v3 CPU running at 2.30GHz.  Each socket has 18 cores, and each core is 2-way 
hyperthreaded, yielding 72 logical CPUs in total.  The system was running Ubuntu 18.04 with a stock 
Linux version 4.15 kernel, and all software was compiled using the provided GCC version 7.3 toolchain
at optimization level ``-O3''.  
64-bit C or C++ code was used for all experiments.  
Factory-provided system defaults were used in all cases, and Turbo mode \cite{turbo} was left enabled.  
In all cases default free-range unbound threads were used.  
SPARC and x86 both provide a strong TSO memory model \cite{Sewell-TSO}.
The atomic fetch-and-add
primitives, \texttt{LOCK:XADD} on x86 and a \texttt{LD;ADD;CAS} loop on SPARC, have fence semantics.
TWA is trivial to implement in C++ with \texttt{std::atomic<>} primitives.  

We implemented all user-mode locks within LD\_PRELOAD interposition
libraries that expose the standard POSIX \texttt{pthread\_\allowbreak{}mutex\_t} programming interface
using the framework from ~\cite{topc15-dice}.  
This allows us to change lock implementations by varying the LD\_PRELOAD environment variable and
without modifying the application code that uses locks.
The C++ \texttt{std::mutex} construct maps directly to \texttt{pthread\_mutex} primitives,
so interposition works for both C and C++ code.
All busy-wait loops used the Intel \texttt{PAUSE} instruction for polite waiting.

We use a 128 byte sector size on Intel processors for alignment to avoid 
false sharing.  The unit of coherence is 64 bytes throughout the cache hierarchy, but 128 bytes
is required because of the adjacent cache line prefetch facility where pairs of lines are automatically 
fetched together.  

\subsection{Sensitivity to Inter-Lock Interference}

\begin{figure}[h!]                                                                    
\includegraphics[width=8.5cm]{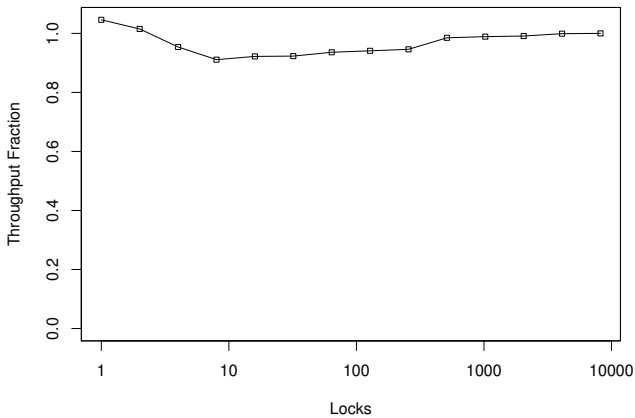}
\caption{Inter-Lock Interference}                                                   
\label{Figure:Interference}                                                                  
\end{figure}

As the waiting array is shared over all locks and threads within an address space, one potential
concern is collisions that might arise when multiple threads are using a large set of
locks.  Near collisions are also of concern as they can cause false sharing within the array.   
To determine TWA's performance sensitity to such effects, we implemented a benchmark program that spawns 64 
concurrent threads.  Each thread loops as follows : randomly pick a lock from a pool of such 
locks; acquire that lock; advance a thread-local pseudo-random number generator 50 steps; release the lock; 
and finally advance that random number generator 100 steps.  At the end of a 10 second measurement interval 
we report the number of lock acquistions.  Each data point is the median of 7 distinct runs.
We report the results in Figure-\ref{Figure:Interference} where X-axis reflects the number of locks in the pool (varying through
powers-of-two between 1 and 8192) and the Y-axis is the number of acquisitions completed by TWA divided
by the number completed by a specialized version of TWA where each lock instance has a private array of 4096 elements.
This fraction reflects the performance drop attributable to inter-lock conflicts and near conflicts in the shared array, where
the modified form of TWA can be seen as an idealized form that has a large per-instance footprint but which is
immune to  inter-lock conflicts.  The worst-case penalty arising from inter-thread interference (the lowest fraction value)
is always under 8\%.

\subsection{MutexBench}

\begin{figure}[h]                                                                    
\includegraphics[width=8.5cm]{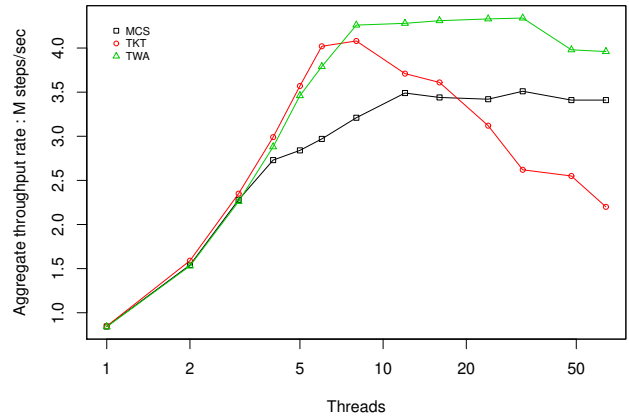}
\caption{MutexBench}                                                   
\label{Figure:MutexBench}                                                                  
\end{figure}

The MutexBench benchmark spawns $T$ concurrent threads. Each thread loops as follows: 
acquire a central lock L; execute a critical section; release L; execute
a non-critical section. At the end of a 10 second measurement interval the benchmark 
reports the total number of aggregate iterations completed by all the threads. 
We show the median of 5 independent runs in Figure-\ref{Figure:MutexBench}.  
The critical section advances a C++ \texttt{std::mt19937} pseudo-random generator (PRNG)
4 steps.  The non-critical section uses that same PRNG to compute a value distributed 
uniformly in $[0,200)$ and then advances the PRNG that many steps.  
For clarity and to convey the maximum amount of information to allow a comparision the algorithms, 
the $X$-axis is offset to the minimum score and the $Y$-axis is logarithmic.
\Invisible{To facilitate comparison; visual comparison; to convey maximum information; for clarity; for density; } 

As seen in the figure, ticket locks performs the best up to 6 threads, with TWA lagging
slightly behind.  As we further increase the thread count, however, ticket locks fail
to scale.  MCS provides stable asymptotic performance that surpasses ticket locks at 24 threads.
TWA manages to always outperform MCS, freeing the developer from making a choice between
MCS locks and ticket locks. 

\begin{figure}[h]                                                                    
\includegraphics[width=8.5cm]{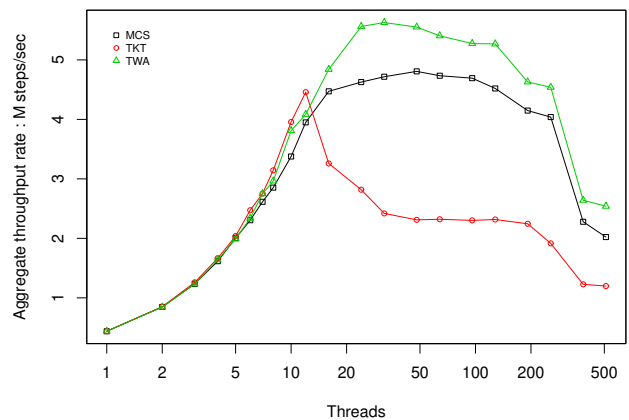}
\caption{MutexBench on Solaris/SPARC 64-bit}                                                   
\label{Figure:MutexBench-sparc64}                                                                  
\end{figure}

To show that our approach is general and portable, we next report MutexBench results on a Sun/Oracle T7-2
\cite{sparc-T7-2}.  The T7-2 has 2 sockets, each socket populated by an M7 SPARC CPU running at 4.13GHz
with 32 cores.  Each core has 8 logical CPUs sharing 2 pipelines.  The system has 512 logical CPUs and 
was running Solaris 11.  64-bit SPARC does not directly support atomic fetch-and-add or swap operations -- 
these are emulated by means of a 64-bit compare-and-swap operator (CASX).  The system uses MOESI cache coherency
instead of the MESIF \cite{MESIF} found in modern Intel-branded processors, allowing more graceful
handling of write sharing.  The graph in Figure-\ref{Figure:MutexBench-sparc64} has the same shape as
found in Figure-\ref{Figure:MutexBench}.  The abrupt performance drop experienced by all locks
starting at 256 threads is caused by competition for pipeline resources.  

\subsection{throw}

\begin{figure}[h]                                                                    
\includegraphics[width=8.5cm]{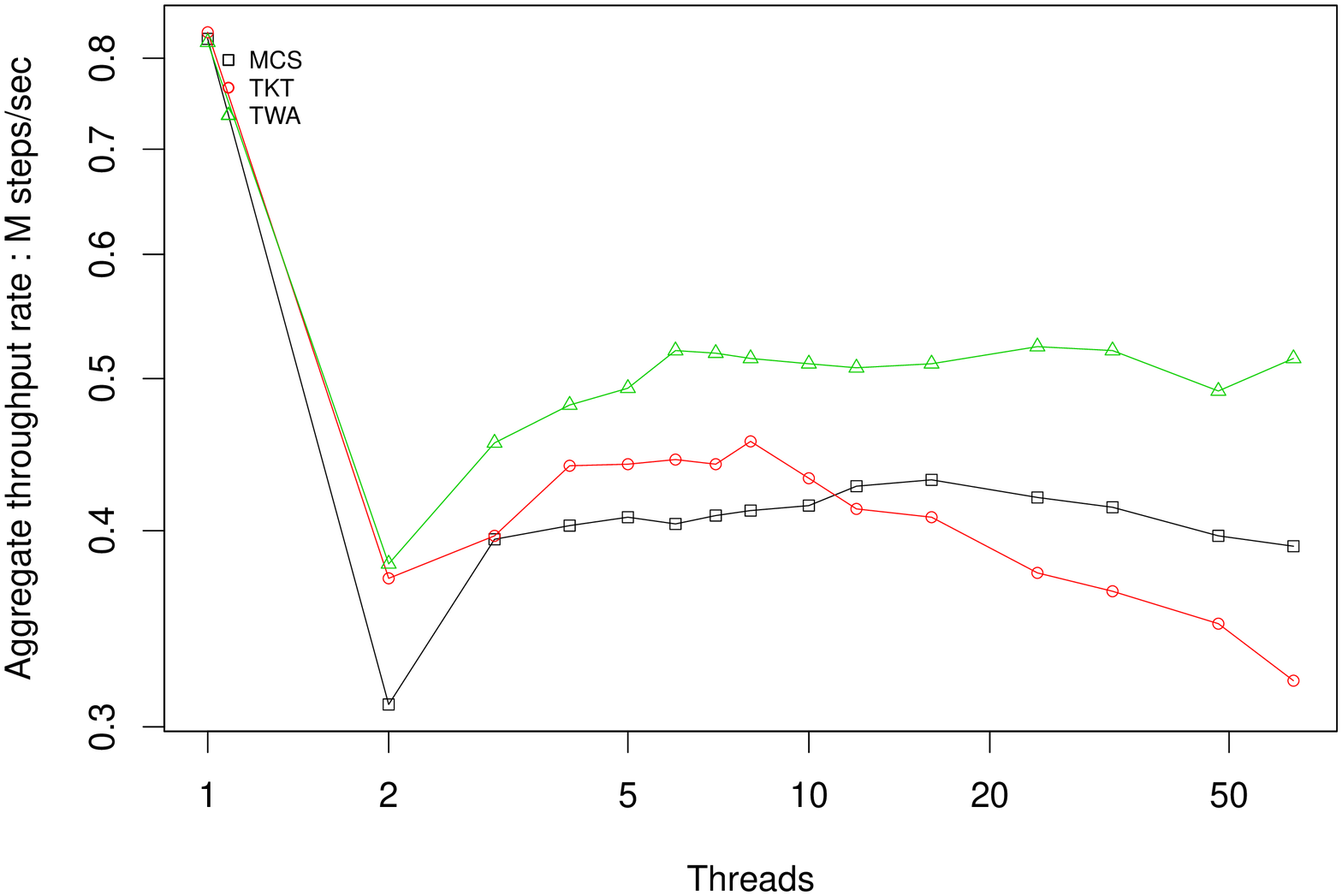}
\caption{throw}                                                   
\label{Figure:throw}                                                                  
\end{figure}

The ``throw'' benchmark launches $T$ threads, each of which loop, executing the following line 
of C++ code: \\ \verb|   try { throw 20 ;} catch (int e) {}|. \\Naively, this construct would be
expected to scale linearly, but the C++ runtime implementation acquires mutexes that protect
the list of dynamically loaded modules and their exception tables
\footnote{The C++ runtime libraries have specialized binding conventions to invoke locking
operators, which are not amenable to normal LD\_PRELOAD interposition.  We instead intercepted the
locking calls via the \texttt{\_rtld\_global} facility.}.  
The problem is long-standing and has proven difficult to fix given the concern that
some applications might have come to depend on the serialization \cite{ThrowCatch}. 
At the end of a 10 second measurement interval the benchmark reports the aggregate number 
of loops executed by all threads.  There is no non-critical section in this benchmark; 
throw-catch operations are performed back-to-back with no intervening delay. 
In Figure-\ref{Figure:throw} we observe that performance drops significantly between 1 and 2 threads.
There is little or no benefit from multiple threads, given that execution is largely serialized,
but coherent communication costs are incurred.  As we increase beyond two threads performance
improves slightly, but never exceeds that observed at one thread.  Beyond 2 threads, the shape of the graph 
recapitulates that seen in MutexBench.  

\subsection{Random Replacement Cache} 

\begin{figure}[h]                                                                    
\includegraphics[width=8.5cm]{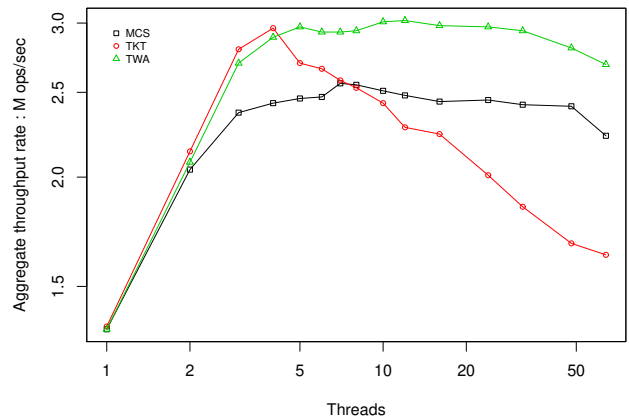}
\caption{Random Replacement Cache}                                                   
\label{Figure:LRUCache}                                                                  
\end{figure}

The ``Random Replacement Cache'' benchmark creates a key-value cache with a random replacement
policy \footnote{We use the cache from \url{https://github.com/ceph/ceph/blob/master/src/common/random\_cache.hpp}
in conjunction with a test harness of our own making}.  All cache operations are protected with a central lock.
Both the keys and values are 32-bit integers and we set values equal to a hash of the key. 
The cache is configured with a capacity limit of 10000 elements.  
The benchmark launches the specified number of concurrent threads, each of which loops, accessing
the cache, and then executing a delay.  At the end of a 10 second measurement interval the benchmark
reports the aggregate throughput rate.  We plot the median of 5 runs for each data point in
Figure-\ref{Figure:LRUCache}.  To emulate locality and key reuse, each thread has a private 
\emph{keyset} of 10 recently used keys.  We pre-populate the keyset with random keys
before the measurement interval, using selection with replacement.
With probability $P=0.9$ a thread picks a random index in its keyset and then uses the corresponding
key for its access.  We use thread-local C++ \texttt{std::mt19937} pseudo-random number generators
with a uniform distribution.  Otherwise, the thread generates a new random key in the range $[0,50000)$,
installs that key into a random slot in the keyset, and then proceeds to access the cache
with that key.  The inter-access delay operation picks a random number in the range $[0,200)$ and then
steps the thread-local random number generator that many times.  

The cache implementation makes frequent use of malloc-free operations.
The default malloc allocator fails to fully scale in this environment and attenuates the 
benefit conferred by improved locks, so we instead used the index-aware allocator from \cite{ismm11-afek}.
This allocator uses its own built-in synchronization primitives instead of \texttt{pthread} operators, 
so LD\_PRELOAD interposition on the \texttt{pthread} mutex primitives has no influence 
on malloc performance.

\subsection{libslock stress\_latency}                     

\begin{figure}[h]                                                                    
\includegraphics[width=8.5cm]{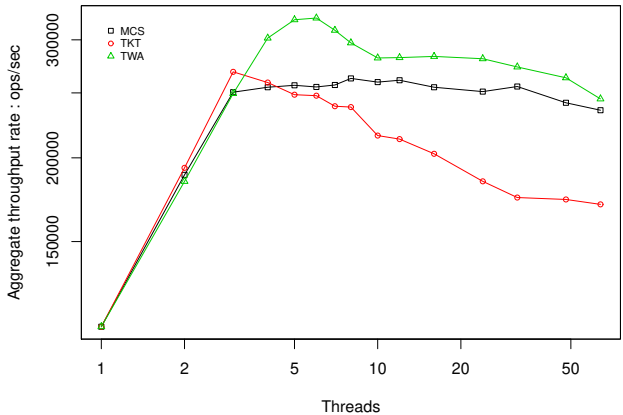}
\caption{libslock stress\_latency}                                                   
\label{Figure:stresslatency}                                                                  
\end{figure}

Figure-\ref{Figure:stresslatency} shows the performance of the ``stress latency''
benchmark from \cite{sosp13-david} \footnote{We use the following command line: ./stress\_latency -l 1 -d
10000 -a 200 -n \emph{threads} -w 1 -c 1 -p 5000.}. 
The benchmark spawns the specified number of threads, which all run concurrently during a 
10 second measurement interval.  Each thread iterates as follows: acquire a central 
lock; execute 200 loops of a delay loop; release the lock; execute 5000 iterations of 
the same delay loop. The benchmark reports the total number of iterations of the outer loop. 

\subsection{LevelDB readrandom}

\begin{figure}[h]                                                                     
\includegraphics[width=8.5cm]{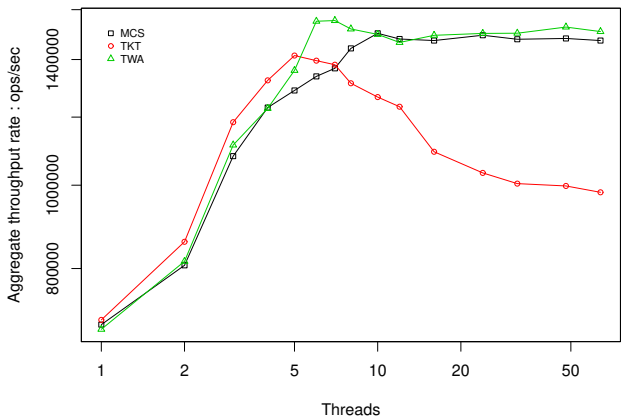}
\caption{LevelDB readrandom}                                                   
\label{Figure:readrandom}                                                                  
\end{figure}

In Figure-\ref{Figure:readrandom}  we used the ``readrandom'' benchmark in LevelDB version 1.20 
database \footnote{\url{leveldb.org}} varying the number of threads and reporting throughput 
from the median of 5 runs of 50 second each.  
Each thread loops, generating random keys and then trying to read the associated value from
the database.  
We first populated a database \footnote{db\_bench \---\---threads=1 
\textendash{}\textendash{}benchmarks=fillseq \---\---db=/tmp/db/}
and then collected data \footnote{db\_bench \---\---threads=\emph{threads}  
\---\---benchmarks=readrandom \\ \---\---use\_existing\_db=1 
\---\---db=/tmp/db/ \---\---duration=50}. 
We made a slight modification to the \texttt{db\_bench} benchmarking 
harness to allow runs with a fixed duration that reported aggregate throughput.  
Ticket locks exhibit a very slight advantage over MCS and TWA at low threads count after
which ticket locks fade and TWA matches or exceeds the performance of MCS.
LevelDB uses coarse-grained locking, protecting the database with a single central mutex: 
\texttt{DBImpl::Mutex}.  Profiling indicates contention on that lock via \texttt{leveldb::DBImpl::Get()}.  

\subsection{LevelDB readwhilewriting}

\begin{figure}[h]                                                                    
\includegraphics[width=8.5cm]{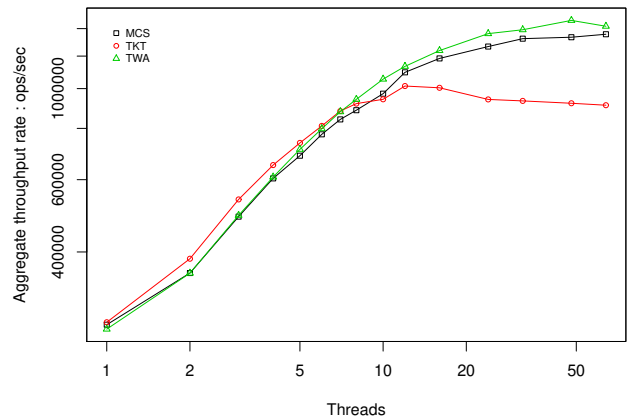}
\caption{LevelDB readwhilewriting}                                                   
\label{Figure:readwhilewriting}                                                                  
\end{figure}

The LevelDB ``readwhilewriting'' benchmark in Figure-\ref{Figure:readwhilewriting} spawns $T-1$ random readers
(identical to the ``readrandom'' threads) and a single writer thread which writes to randomly selected keys 
\footnote{db\_bench \---\---benchmarks=readwhilewriting \---\---threads=\emph{threads} \---\---cache\_size=50000 \---\---num=100000 \---\---duration=50}.  The benchmark reports the aggregate throughput completed in a 50 
second measurement interval.  Each data point is taken as median of 5 distinct runs.   
The same lock in ``readrandom'' is the source of contention in this benchmark.  

\subsection{RocksDB readwhilewriting}

\begin{figure}[h]                                                                    
\includegraphics[width=8.5cm]{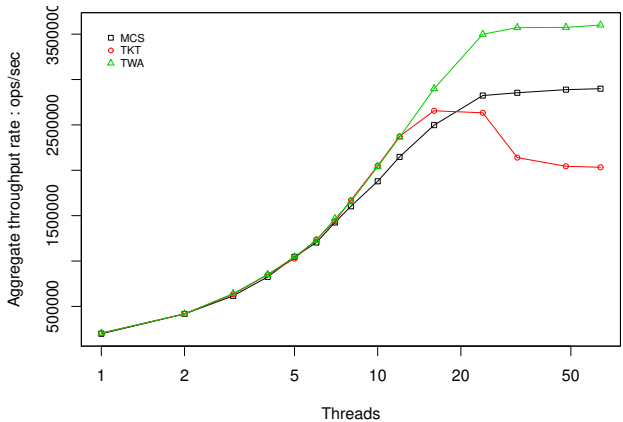}
\caption{RocksDB readwhilewriting}                                                   
\label{Figure:rocksdb}                                                                  
\end{figure}

We next present results in Figure-\ref{Figure:rocksdb} from the RocksDB \footnote{\url{rocksdb.org}} 
version 5.14.2 database running their variant of the ``readwhitewriting'' benchmark.  
The benchmark is similar to the form found in LevelDB, above, but the underlying database 
allows more concurrency and avoids the use of a single central lock.  
We intentionally use a command-line configured to stress the locks that protect the sharded 
LRU cache, causing contention in \texttt{LRUShard::lookup()}
\footnote{db\_bench \---\---duration=200 \---\---threads=\emph{threads} \\ \---\---benchmarks=readwhilewriting
\---\---compression\_type=none \\ \---\---mmap\_read=1 \---\---mmap\_write=1 
\---\---cache\_size=100000 \\ \---\---cache\_numshardbits=0 
\---\---sync=0 \---\---verify\_checksum=0}. 

\subsection{Linux kernel \texttt{locktorture}} 

We ported TWA into the Linux kernel environment and evaluated its performance with the 
\texttt{locktorture} benchmark \footnote{\url{https://www.kernel.org/doc/Documentation/locking/locktorture.txt}}. 
\texttt{Locktorture} is distributed as a part of the Linux kernel.  It is implemented as 
a loadable kernel module, and according to its documentation, ``runs torture tests on
core kernel locking primitives'', including \emph{qspinlock}, the kernel spin lock. 
It creates a given number of threads that repeatedly acquire and release the lock, with occasional short
delays (citing the comment in the source code, ''to emulate likely code'') and occasional long delays 
(''to force massive contention'') inside the critical section.  At the end of the measurement interval, it
reports the overall throughput (lock acquisitions) completed by the threads.  We used locktorture to
compare TWA, classic ticket locks, and the default kernel qspinlock.  

The Linux \emph{qspinlock} construct \cite{linux-locks,linux-page-struct,Long13} is a compact 32-bit lock, even
on 64-bit architectures.  The low-order bits of the lock word constititue a simple test-and-set lock while
the upper bits encode the tail of an MCS chain.  The result is a hybrid of MCS and 
test-and-set\footnote{\url{https://github.com/torvalds/linux/blob/master/kernel/locking/qspinlock.c}}. 
In order to fit into a 32-bit work -- a critical requirement -- the chain is 
formed by logical CPU identifiers instead of traditional MCS queue node pointers.  
Arriving threads attempt to acquire the test-and-set lock embedded in the low order bits 
of the lock word.  This attempt fails if the test-and-set lock is held or of the MCS chain is populated.  
If successful, they enter the critical section, otherwise they join the MCS chain embedded
in the upper bits of the lock word.  When a thread becomes an owner of the MCS lock, it can wait for the
test-and-set lock to become clear, at which point it claims the test-and-set lock, releases the MCS lock, 
and then enters the critical section.  The MCS aspect of qspinlock is used only when there is contention. 
The unlock operator simply clears the test-and-set lock.  The MCS lock is never held over the critical section,
but only during contended acquistion.  Only the owner of the MCS lock spins on the test-and-set lock,
reducing coherence traffic \footnote{This provides a LOITER-style \cite{eurosys17-dice} lock with the \emph{outer lock} 
consisting of a test-and-set lock and the \emph{inner lock} consisting of the MCS lock, with both locks embedded 
in the same 32-bit word.}.  Qspinlock is strictly FIFO.  While the technique employs local spinning on the MCS chain, 
unlike traditional MCS, arriving and departing threads will both update the common lock word, 
increasing coherence traffic and degrading performance relative to classic MCS.   
Qspinlock incorporates an additional optimization where the first contending thread spins on the test-and-set lock 
instead of using the MCS path.  Traditional MCS does not fit well in the Linux kernel as (a) the contraint 
that a low-level spin lock instance be only 32-bits is a firm requirement, and (b) the lock-unlock 
API does not provide a convenient way to pass the owner's MCS queue node address from lock to unlock.  
We note that qspinlocks replaced classic ticket locks as the kernel's primary low-level spin lock 
mechanism in 2014, and ticket locks replaced test-and-set locks, which are unfair and allow unbounded bypass, 
in 2008 \cite{ticket-spinlocks}.   

\renewenvironment{quotation}%
  {\list{}{\leftmargin=0.16in\rightmargin=0.16in}\item[]}%
  {\endlist}

Regarding lock instance size, Bueso \cite{cacm15-Bueso} notes:
\begin{quotation} 
\textbf{Lock overhead}. This is the resource cost of using a particular lock in terms of both size and latency. 
Locks embedded in data structures, for example, will bloat that type. Larger structure sizes mean more CPU cache 
and memory footprint. Thus, size is an important factor when a structure becomes frequently used throughout the system. 
Implementers also need to consider lock overhead when enlarging a lock type, after some nontrivial modification; this 
can lead to performance issues in unexpected places. For example, Linux kernel file-system and memory-management 
developers must take particular care of the size of VFS struct inode (index node) and struct page, optimizing as 
much as possible. These data structures represent, respectively, information about each file on the system and each 
of the physical page frames. As such, the more files or memory present, the more instances of these structures 
are handled by the kernel. It is not uncommon to see machines with tens of millions of cached inodes, so increasing 
the size of the inode by 4 percent is significant. That's enough to go from having a well-balanced workload to not 
being able to fit the working set of inodes in memory. Implementers must always keep in mind the size of the locking primitives.
\end{quotation} 
Another example where the size of the lock is important is in concurrent data structures, such as linked lists or 
binary search trees, that use a lock per node or entry \cite{ppopp10-Bronson, ppopp12-Crain, opodis05-Heller}. 
As Bronson at el. observe, when a scalable lock is striped across multiple cache lines to 
avoid contention in the coherence fabric, it is ``prohibitively 
expensive to store a separate lock per node''\cite{ppopp10-Bronson}. 

\Invisible{qspinlock is hybrid test-and-set and MCS lock -- compound ``LOITER'' lock} 

In Table-\ref{Complexity} we report the cyclomatic complexity \cite{mccabe-cyclomatic}
and N-Path complexity \cite{Nejmeh-NPath,Bang-FSE15} measures -- derived from the complexity of the
control flow graph -- for the lock and unlock methods
for classic ticket locks (TKT), the kernel qspinlock primitive
\footnote{\url{https://github.com/torvalds/linux/blob/master/kernel/locking/qspinlock.c}} and TWA.  
To facilitate a fairer comparison, we removed performance monitoring and debugging facilities from qspinlock
before running \texttt{oclint} \footnote{\url{https://github.com/oclint/oclint}} to compute the complexity.
In addition, the values reported above reflect only the top-level qspinlock acquistion method, and
does not include the complexity of helper methods that it calls.  The only helper methods used
by TWA and ticket locks are atomic operators and the PAUSE operator, which reduce to just one instruction
with no additional complexity.  All three lock algorithms have cyclomatic and N-Path complexity of just 1 for 
the unlock operation.  As can be seen in the table, TWA's lock operator is far less complex than
is the qspinlock acquisition method.

\Invisible{NPath complexity is exponential function of cyclomatic complexity} 

\Invisible{Implicitly, we're suggesting that complexity measure correlate with understandability
and bug rates -- correctness.  Code with higher complexithy is more apt to contain latent flaws
and is harder to analyze and understand.  We use complexity as a proxy for correctness in the
absense of formal proofs.}  


\begin{table} [h]
\centering
\begin{tabular}{lcccc}
\toprule
\multicolumn{1}{l}{} &
\multicolumn{2}{c}{NPath}       &
\multicolumn{2}{c}{Cyclomatic}  \\
\cmidrule(lr){2-3}
\cmidrule(lr){4-5}

&
\multicolumn{1}{c}{Lock} &
\multicolumn{1}{c}{Unlock} &
\multicolumn{1}{c}{Lock} &
\multicolumn{1}{c}{Unlock} \\
\midrule

Ticket Lock   & 2    & 1 & 2  & 1 \\
QSpinLock     & 4320 & 1 & 18 & 1 \\
TWA           & 28   & 1 & 6  & 1 \\

\midrule[\heavyrulewidth]
\bottomrule
\end{tabular}%
\caption{Complexity Measures}\label{Complexity}
\end{table}

\begin{figure}[h]                                                                    
\includegraphics[width=8.5cm]{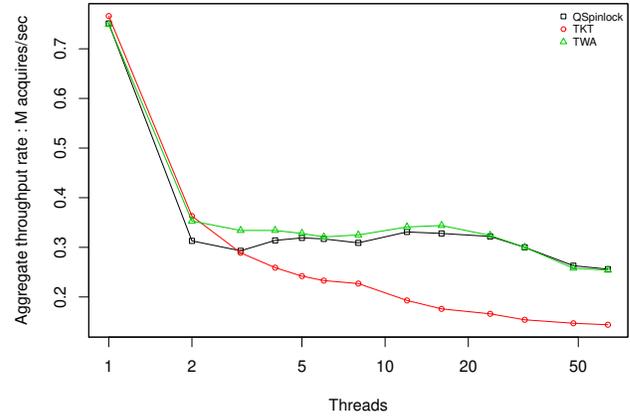}
\caption{LockTorture : High Contention}                                                   
\label{Figure:LockTorture-HighContention}                                                                  
\end{figure}

\begin{figure}[h]                                                                    
\includegraphics[width=8.5cm]{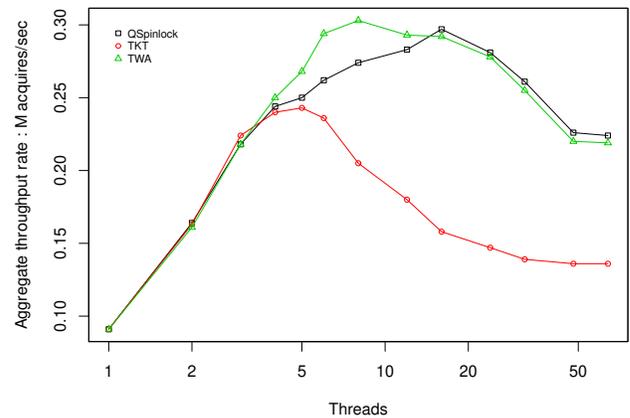}
\caption{LockTorture : Moderate Contention}                                                   
\label{Figure:LockTorture-ModerateContention}                                                                  
\end{figure}

The average critical section duration used by \texttt{locktorture} is a function of the 
number of concurrent threads.  In order to use the benchmark to measure and report scalability, we augmented it to
parameterize the critical and non-critical section durations, which are expressed as steps of the thread-local
pseudo-random number generator provided in the \texttt{locktorture} infrastructure.  We used 20 steps 
for the critical section.  Each execution of the non-critical section computes a uniformly random distributed
number in $[0-N]$ and then steps the local random number generator that many iterations.  
At the end of a run (lasting 30 seconds in our case), the total number of lock operations performed by all 
threads is reported.  We report the median of 7 such runs.    
Figure-\ref{Figure:LockTorture-HighContention} uses $N=20$ and Figure-\ref{Figure:LockTorture-ModerateContention}
uses $N=400$.  

For the kernel versions of ticket locks and TWA we reduced the \texttt{ticket} and \texttt{grant} fields 
to 16 bits to allow the composite lock word to fit in 32-bits, imposing a constraint that the maximum number 
of waiting threads never exceeds 65535.  Qspinlocks operate with preemption disabled, so the constraint devolves to 
not having more than 65535 \emph{processors} waiting on a given lock at any one time.
By default, the maximum number of CPUs allowed by the Linux kernel on x86 architectures is 4096, satisfying
the constraint. 

As we can see in Figures \ref{Figure:LockTorture-HighContention} and \ref{Figure:LockTorture-ModerateContention},
classic ticket locks perform well at low conconcurrency but fade as the number of threads increases.
TWA and and QSpinLock provide fairly similar performance, although TWA is far simpler.  

\section{Conclusion}

while our approach is deterministic
we note that performance under TWA can be influenced by the activities of other unrelated threads and locks
by means of collisions in the shared waiting array, potentially reducing predictability.  Other
shared resources incur the same risk. Examples include (a) competition for occupancy of shared hardware caches
and (b) collisions in the Linux \emph{futex} hash table, where lock addresses map to hash chains of blocked threads. 

\Invisible{Any structure that uses a lock to protect hash chains}  

TWA is a straightforward extension to classic ticket locks, providing the best performance
properties of ticket locks and MCS locks.  Like ticket locks, it is simple, compact, and has a fixed memory footprint. 
The key benefit conferred by TWA arises from
improved transfer of ownership (handover) in the unlock path, by reducing the number
of threads spinning on the \texttt{grant} field at any given time.  
Even though TWA increases the overall path length in the unlock operation, adding an atomic
fetch-and-increment operation compared to the classic ticket lock, it decreases the 
effective critical path duration for contended handover.

\Invisible{In the Appendix we identify a number of variations on the basic TWA algorithm that we plan to
explore in the future.}

We plan to explore long-term waiting in the Linux kernel via the MONITOR-MWAIT construct or by
means of the kernel \emph{futex} \cite{futex} mechanism. 

\AtFoot{An extended version of this paper is available at \url{https://arxiv.org/abs/1810.01573}} 



\begin{acks}                            

We thank Shady Issa for reviewing drafts and helpful comments.


\end{acks}

\bibliography{TWA.bib}

\section{Appendix : Algorithmic Variations}

\Boldly{TKT-Dual} An interesting variation on TWA is to forgo the waiting array and simply augment the ticket
lock structure to use \emph{two} grant fields, one for short-term waiting, for the 
immediate successor and perhaps a small number of other threads ``near'' the front of 
conceptual queue, and a second grant field for long-term waiting. 
To reduce coherence traffic we isolated the grant field to be sole occupant of a cache sector, 
increasing the lock size.
The unlock operator first advances the short-term grant field, and, as needed, may 
advance the long-term grant field to shift one or more threads from long-term to 
short-term waiting, constraining the number of short-term waiters and accelerating handover.
We refer to this form as \emph{TKT-Dual} given the dual encoding of the grant field. 
Initial experiments with this form show promise, yielding results better 
than that of the baseline ticket lock, although lagging slightly behind TWA.  
The increased size means this form can not be used in the linux kernel as a replacement for the
qspinlock algorithm.

\Boldly{TWA-Staged} We are also exploring variations of TWA where the fast uncontended path for both 
lock and unlock operations would be identical to that of normal ticket locks,
specifically avoiding accesses to the waiting array in the unlock operator. 
Briefly, we divide waiting threads into 3 groups:
(A) Those that are 2 or more elements away from the front of the conceptual queue.               
These ``long-term'' threads wait via the waiting array in the usual TWA fashion.
(B) The thread that is 2 away from the head. 
This ``transitional'' thread busy-waits on the global \texttt{grant} field.  
(C) The thread that is 1 away from the head.  This is the immediate successor, 
and also busy-waits on the global \texttt{grant} field.  
Assume we have a non-trivial set of waiting threads.  Incrementing \texttt{grant} in the unlock
operator passes ownership to the thread \emph{T1} in (C) state above.  
\emph{T1} exits (C) state and becomes the owner. 
Thread \emph{T2} in (B) state also observes the change in the \texttt{grant} field, at which point 
it increments the waiting array slot associated with the next ticket value -- 
the ticket value one after it's assigned ticket -- to transfer a thread from 
(A) state to take \emph{T2's} place as the (B) thread.  
\emph{T2} then shifts from (B) to (C) state.  
The downside to this approach is that we have 2 threads directly busy-waiting on 
\texttt{grant}, the (B) and (C) threads, whereas in our other TWA variants we at 
most have one thread busy-waiting on \texttt{grant}.  
The upside is that unlock operator does not access the waiting array, and 
all the waiting array accesses -- both loads and fetch-and-add operations -- are 
performed by waiting threads.  As noted above, the unlock operator simply increments the grant field.
This approach leverages those waiting threads to help drive through 
the (A)$\rightarrow$(B)$\rightarrow$(C) transitions, reducing the path length of the unlock operation.
Relative to classic ticket locks, all additional code is encapsulate in the locking slow path.
In the case of uncontended operations, the array is never accessed and the path is the same as 
classic ticket locks, improving uncontended unlock latency.   

\Boldly{TWA-ID} We note that we can replace the atomic increment of the waiting array element in the unlock path
with a simple store of 0 to that location by changing the waiting array elements from
counters to unique thread identity references.  Threads arriving in the long-term waiting
state will write their temporally unique non-zero identity value (which can be as simple as 
the address of the stack pointer) into
the waiting array, recheck the \texttt{grant} value and then busy-wait while the
waiting array elements remains equal to the thread identity value they just stored.  
When the element changes, the thread shifts to classic short-term waiting.  
While this approach eliminates the atomic fetch-and-add in the unlock path, it also
increases write traffic into the shared array, as threads entering the long-term write
phase must store their unique identity.

\Boldly{TWA-Quantized} To reduce the impact on the \texttt{unlock} path we can arrange
for the \texttt{unlock} operator to release a quanta or gang of $N$ threads -- 
based on the ticket lock values -- into global spinning mode every $Nth$ unlock operation.   
This reduces the average cost of an \texttt{unlock} operation, but
increases the number of threads waiting via global spinning. 

\Boldly{3-Stage} 
A key observation in the design of TWA is importance of hand-over efficiency, improved by
reducing the number of threads concurrently busy-waiting on the \texttt{grant} field.  
This insight informs the design of other locks that perform gracefully under contention.  
One such variation is a \emph{3-Stage} mutex $M$ which has 3 sub-locks : $A$, $B$ and $C$.  
$A$ would typically by a FIFO lock such as MCS or ticket 
$B$ and $C$ could be simple test-and-set locks.  
To acquire $M$ a thread uses the following protocol : acquire $A$; acquire $B$; release $A$; acquire $C$; release $B$.
To release $M$ a thread simply releases $C$.  
The thread that holds $C$ also holds $M$ itself.   
(For efficiency, we might shift the release of $B$ to unlock path).  

\Invisible{
* Handoff; handover
* The basic operating theme is to get the expensive handoffs out of the critical paths.   
} 

This approach, while seemingly complicated,  confers a number of advantages.   
If we have $N$ waiting threads, then $N-2$ will typically wait on $A$, $1$ on $B$ and $1$ on $C$.  
When we release $M$, by releasing $C$, there is at most one thread waiting on $C$, so handover costs 
arising from coherence traffic are minimized.  Once a thread has acquired $C$ it still needs to release $B$, 
and this occurs within the critical section of $M$.  
But there is at most one waiter on $B$, so the cost to release $B$ isn’t excessive.   
A relatively large handover cost may be incurred when we release $A$, but that latency overlaps 
with execution of the critical section and does not manifest as reduced throughput.  
The latency associated with the release of $A$ is subsumed into the waiting period for $M$ and moved out of the
critical path, improving throughput and scalability.   
Note that $M$ is FIFO if $A$ is FIFO. 
$A$ protects the acquisition of $B$, so as an optional optimization we can avoid atomics to acquire $B$.  
And $B$ likewise protects the acquisition of $C$.  

To reduce the path length in the uncontended case, we can add a fast-path where
arriving threads \emph{trylock} directly on $C$, and, if successful can avoid $A$ and $B$.  
In this case, however, we would need to use atomics to acquire $C$ in the slow path.  

\Invisible{And or course we can add a fast path where arriving threads trylock directly on C and skip all 
the A+B nonsense that exists to mitigate contention.  Once we add a fast path with bypass we 
probably need to add some anti-starvation mechanism, but that’s not hard.  Basically, if the 
owner of B waits too long it can become impatient and require direct handoff the next time C 
is dropped.   And we need to use atomics to acquire C.
In the simple formulation without a trylock-based fast path, B and C don’t need atomics, as there would be
at most one thread trying to acquire those sub-locks at any given time.  So we can get by with loads and stores. } 

\Invisible{The fetch-and-increment of the waiting array element in the unlock operator 
can be avoided as follows.  The waiting array contains unique thread identities instead of 
notification counters.  Threads start a long-term waiting phase by storing 
their non-zero identity into the array and then recheck the value of \texttt{grant}.
They then busy-wait while the array element remains equal to their own identity.
The unlock operator simply stores 0 into the waiting array instead of incrementing the location.} 

\Invisible{ We also intend to look into ways to avoid the precautionary 
increment in the unlock path in the waiting array if there are no waiters, 
or no long-term waiters.  We hope to investigate the use of the external waiting array
for other locks, such as the ``tidex'' algorithm \cite{ppopp17-Ramalhete}.}
  
\Invisible{We also plan on kernel-level experiments to determine if TWA might be a
viable replacement for the Linux kernel's existing \emph{qspinlock} construct.} 

Finally, we believe that replacing the waiting array elements with pointers to
chains of waiting threads may have benefit.  Briefly, each long-term waiting thread 
would have an on-stack MCS-like queue node that it would push into the appropriate chain
in the waiting array, and then use local spinning on a field within that node.
Notification of long-term waiters causes the chain to be detached via an 
atomic \texttt{SWAP} instruction and all the elements are updated to reflect that 
they should reevaluate the \texttt{grant} field.  In the case of collisions, 
waiting threads may need to re-enqueue on the chain.  
This design recapitulates much of the Linux kernel ``futex'' mechanism.  

\Invisible{For environments that already use ticket locks, and where the size of the ticket lock is 
``baked'' into binaries, TWA may serve as a drop-in replacement if interposition is available.  
The same is not true TKTDual where an additional grant field must be added to the lock stru`cture.} 

\Invisible{ Recall that for TWA, the locking fast path is exactly the same as TKT, but the unlock 
operator adds that precautionary increment into the waiting array above and beyond the normal unlock code, 
so we’d like to get rid of that increment if it’s not needed.  

Under little or no contention, I can speed up unlock by trying to undo the increment of ticket with a CAS.  
That is, at unlock time we try to revert the ticket value with a CAS instead of incrementing grant.  
If the CAS succeeds, then we can skip bumping the notification indicator in the waiting array, 
and as bonus, we don’t cyclically touch all the lines under the waiting array with those 
precautionary increments, in turn decreasing D-cache pressures.   This is slightly faster than TWA at 
low/no contention, but falls about midway between TKT and TWA at higher levels of contention.  

Another idea to eliminate the increment into the waiting array in unlock is as follows.  
We ``stage'' the waiting threads, so the immediate successor, as usual, waits directly on the grant field.  
We’ll call the immediate successor S1.   (S1 identifies a specific role, not a thread).  
That threads successor, which we’ll S2, also waits directly on grant.   Other waiters deeper in the logical 
queue wait via the long-term waiting array.   So we have 2 threads busy-waiting on grant, which isn’t ideal.  
The subsequent unlock operation will pass ownership to S1.  S2 also notices that grant changed, and takes 
on and assumes the role of the immediate successor S1.  As S2 transitions itself to the S1 position, it 
performs the fetch-and-add into the waiting array to notify its successor to exit long-term waiting and 
to take over the role of S2.   So we’ve eliminated the increment into the waiting array from the unlock path, 
and made the waiting threads take over that responsibility.   The performance is just a tiny bit better than 
TWA under light/no contention.  And under high contention performance falls between classic TKT and TWA.   
This idea sounds good in principle, but it doesn’t really work.  I think the issue is that there’s still a 
real jump in the ``invalidation diameter'' between having 1 and 2 threads busy waiting on the location.   
} 

\section{Appendix: Maximum Ideal Scalability} 

\begin{figure}[h]
\includegraphics[width=8.5cm]{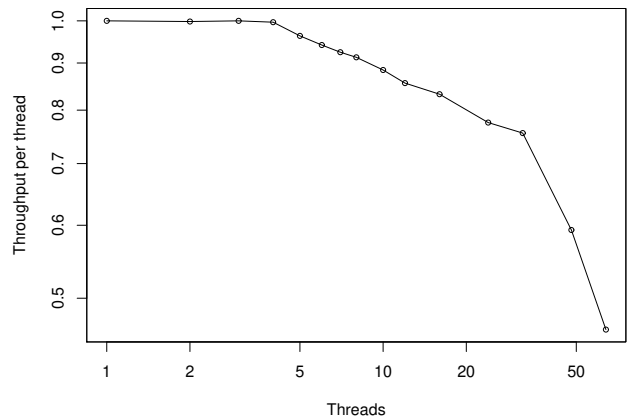}
\caption{Maximum Ideal Scalability}
\label{Figure:MaximumIdealScalability}
\end{figure}

We note that our system-under-test does not exhibit ideal linear scalability as we increase the number of threads.
To demonstrate this effect we use a \texttt{IdealScalabilty} microbenchmark which spaws $T$ concurrent threads,
each of which loops advancing a thread-local \texttt{std::mt19933} pseudo-random number generator.  
There is no sharing, no communication, and no waiting; the threads are completely independent.  
At the end of a 30 second measurement interval the benchmark tallies and reports the aggregate number of 
random number steps completed.  We report the median of 5 such runs for each data point.  
In Figure~\ref{Figure:MaximumIdealScalability} we vary $T$ on the X-axis, and on the Y-axis we report
the normalized throughput -- throughput divided by the throughtput at $T=1$ -- divided by the number 
of participating threads $T$.  This yields a fraction which reflects the slow-down caused by fraternal 
interference for shared hardware compute resources, such as caches, cores, pipelines, DRAM channels, etc.  
Specifically, we observe that the progress rate of each thread is impeded by the concurrent execution of 
unrelated threads.  

\end{document}